\documentclass[9pt]{amsart}

\usepackage{amsmath,amsfonts,amssymb}
\usepackage[dvips]{graphicx}

\usepackage[authoryear,sort&compress]{natbib}

\usepackage{array,delarray}

\newtheorem{theorem}{Theorem}[section]   % Numbered within each section

\newtheorem{corollary}[theorem]{Corollary}     % Numbered along with thm
\newtheorem{proposition}[theorem]{Proposition}  % Numbered along with thm

\theoremstyle{definition}
\theoremstyle{remark}
\newtheorem{example}[theorem]{Example}        % Numbered along with thm

\numberwithin{equation}{section}     % Number equations within sections
\begin{document}

\title{Reduction of Boolean Networks}

\author{Alan Veliz-Cuba}
\address{Department of Mathematics, Virginia Tech, Blacksburg,
VA; Virginia Bioinformatics Institute, Blacksburg, VA}
\email{alanavc@vbi.vt.edu}

\begin{abstract}
Boolean networks have been successfully used in modelling gene
regulatory networks. In this paper we propose a reduction method
that reduces the complexity of a Boolean network but keeps dynamical
properties and topological features and hence it makes the analysis
easier; as a result, it allows for a better understanding of the
role of network topology on the dynamics. In particular, we use the
reduction method to study steady states of Boolean models.
\end{abstract}

\maketitle

%%%%%%%%%%%%%%%%%%%%%%%
\section{Introduction}
%%%%%%%%%%%%%%%%%%%%%%%

Boolean networks have been successfully used in modelling gene
regulatory networks such as the \textit{Drosophila} segment polarity
network \cite{albert}, the yeast cell-cycle network \cite{li} and
the Th regulatory network \cite{mendoza}. Boolean networks provide a
nice theoretical framework that allows for simulation, control
theory and reverse engineering.

However, their analysis is not a trivial task. For example, the
problem of finding steady states has been shown to be NP-complete
\cite{Quianchuan}. One way to overcome these kind of problems is to
develop mathematical and computational tools
\cite{ginsim,JLSS,dvd,RemmyRuet}. While these tools allow to answer
some questions, such as what the steady states are, it is often not
intuitive why such answers were obtained. Another way is to reduce
the network to one that has less complexity while keeping the main
features; the reduced network is easier to analyze and can not only
help to answer questions, but also to give insight of why such
answers were obtained. This in turn, provides a better understanding
of the problem that is being studied.

In this paper a reduction method for Boolean networks is proposed; a
preliminary formulation of this method was provided in \cite{SV}.
The reduction method reduces the complexity of Boolean networks
making the analysis easier and also elucidates the role of network
topology in dynamics. We will focus on the existence, number and
type of steady states. A similar reduction method for logical models
has been proposed in \cite{LMred}.

This paper is organized as follows, in Section \ref{sec-red} we
present the reduction method; properties are presented in Section
\ref{sec-pro}. Section \ref{sec-app} includes an application of the
reduction method. We close with a discussion in Section
\ref{sec-disc}.

%%%%%%%%%%%%%%%%%%%%%%%%%%%%%%%%%%%%%
\section{Reduction Method}
\label{sec-red}
%%%%%%%%%%%%%%%%%%%%%%%%%%%%%%%%%%%%%

%%%%%%%%%%%%%%%%%%%%%%%%%%%%%%%%%%%%%
\subsection{Reduction Steps}
\label{sec-red-steps}
%%%%%%%%%%%%%%%%%%%%%%%%%%%%%%%%%%%%%

We now provide the reduction steps to reduce a Boolean network and
its corresponding  wiring diagram. The idea behind the reduction
method is simple: the wiring diagram and Boolean functions should
reflect direct regulation and hence nonfunctional edges and
variables should be removed; on the other hand, vertices can be
deleted, without losing important information, by allowing its
functionality to be ``inherited'' to other variables.

\begin{enumerate}
    \item We simplify the Boolean functions and wiring diagram:
    \begin{enumerate}
    \item Reduce Boolean expressions using Boolean algebra. This
    will delete variables that are not functional.
    \item Delete edges that do not correspond to Boolean
    expressions. That is, we delete edges that are non functional.
    \end{enumerate}

    \item We delete vertices with no self loop, that is, vertices whose Boolean
    function does not depend on it. Let $x_i$ be a vertex such that
    $f_{x_i}$ does not depend on $x_i$.
    \begin{enumerate}
    \item For all vertices $x_i\rightarrow y$, that is, for all vertices whose
    Boolean function depends on $x_i$, replace the Boolean function for $y$,
    $f_y(x_1,\ldots,x_i,\ldots,x_k)$, by $f_y(x_1,\ldots,f_{x_i},\ldots,x_k)$.
    \item Replace edges $y \rightarrow x_i\rightarrow z$ by $y\rightarrow z$ and delete
    $x_i$ (and edges from/to $x_i$)
    \end{enumerate}
\end{enumerate}

We will see that these steps give rise to a reduced network that
keeps features of the wiring diagram and dynamical properties of the
original network.

%%%%%%%%%%%%%%%%%%%%%%%%%%%%%%%%%%%%%
\subsection{Reduction Algorithm}
\label{sec-red-alg}
%%%%%%%%%%%%%%%%%%%%%%%%%%%%%%%%%%%%%

We present an algorithm to simplify Boolean functions and their
wiring diagram (S); and an algorithm to eliminate a vertex $x$ (R)
%(see \si)
.

%%%%%%%%%%%%%%%%%%%%%%%%%%%%%%%%%%%%%
\subsubsection{Algorithm S} \label{sec-red-alg-S}
%%%%%%%%%%%%%%%%%%%%%%%%%%%%%%%%%%%%%
Input: $f=(f_1,\ldots,f_n)$ and $A$.

\begin{enumerate}
\item For $i=1,\ldots,n$:

\item Simplify $f_i$ using Boolean algebra

\item Construct $A$ corresponding to variables appearing in $f_i$

\end{enumerate}

Output: (Simplified) $f=(f_1,\ldots,f_n)$ and $A$.

\begin{figure}[here]
\centerline{ \hbox{ \framebox{
\includegraphics[width=0.6\textwidth]{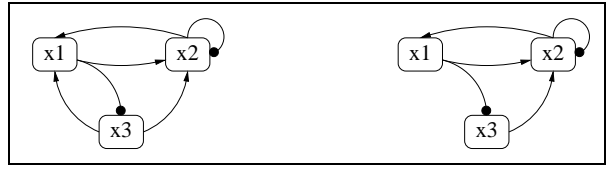}}}
 }\caption{Wiring diagram of $f$ before (left) and after (right) using algorithm S.
 Arrows indicate positive
  paths and circles indicate negative paths.}
 \label{figure_S}
\end{figure}

\begin{example}
Consider the Boolean network given by $f=(f_1,f_2,f_3)=((x_2\wedge
x_3)\vee x_2,(x_1\wedge x_3)\vee \neg x_2,\neg x_1)$ with wiring
diagram given in Figure \ref{figure_S}. Algorithm S gives as an
output %(see \si for details)
$f=(f_1,f_2,f_3)=(x_2,(x_1\wedge x_3)\vee \neg x_2,\neg x_1)$ with
wiring diagram given in Figure \ref{figure_S}. We can clearly see
that Algorithm S detected that the although the variable $x_3$
appears in $f_1$, it is not functional; hence $x_3$ is removed from
$f_1$ (using Boolean algebra) and the edge $x_3\rightarrow x_1$ is
deleted as well. That is, after using S, we obtain an accurate
representation of $f$ and its wiring diagram.
\end{example}

If we say that a variable $z$ depends on $x$, we mean that $f_z$
depends on $x$. Also, if we write $f_i=f_i(x_{i_1},\ldots,x_{i_k})$,
we are saying that we are considering the function $f_i$ in terms of
$x_{i_1},\ldots,x_{i_k}$, even if $f_i$ does not depend on some of
the $x_{i_j}$'s; however, it does mean that $f_i$ does not depend on
the other variables. This convention will make some of the
definitions and proofs simpler.

Let $f=(f_1,\ldots,f_n)$ be a Boolean network with wiring diagram
given by the adjacency matrix, $A$. We show the algorithm to reduce
$f$ by eliminating a vertex $x$ that does not have a self loop. The
algorithm to eliminate $x$ is as follows:

%%%%%%%%%%%%%%%%%%%%%%%%%%%%%%%%%%%%%
\subsubsection{Algorithm R} \label{sec-red-alg-R}
%%%%%%%%%%%%%%%%%%%%%%%%%%%%%%%%%%%%%

Input: $f=(f_1,\ldots,f_n)$, $A$ and $x$ that does not depend on
itself.

\begin{enumerate}
\item Find the variables that depend on $x$: $z_1,\ldots,z_r$

\item For $i=1,2,\ldots,\ldots,r$ \\
Replace $f_{z_i}=f_{z_i}(x,\ldots)$ by $f_{z_i}=f_{z_i}(f_x,\ldots)$

\item Let $f^{[x]}=(f_1,\ldots,\check{f_x},\ldots,f_n)$
(where $\check{f_x}$ means that $f_x$ is omitted) and simplify it
using S.

\item Let $A^{[x]}$=adjacency matrix of the wiring diagram of $f^{[x]}$

\end{enumerate}

Output: $f^{[x]}$ and $A^{[x]}$.

\begin{figure}[here]
\centerline{ \hbox{ \framebox{
\includegraphics[width=0.6\textwidth]{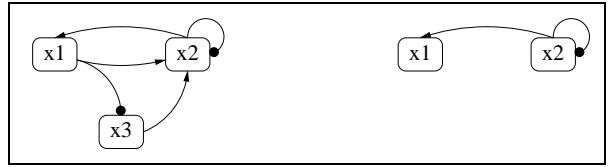}}}
 }\caption{Wiring diagram of $f$ before (left) and after (right) using
  algorithm R to eliminate $x_3$.}
 \label{figure_R}
\end{figure}

\begin{example}
Consider the Boolean network given by
$f=(f_1,f_2,f_3)=(x_2,(x_1\wedge x_3)\vee \neg x_2,\neg x_1)$ with
wiring diagram given in Figure \ref{figure_R}. After using algorithm
R we obtain the network $f^{[x_3]}=h=(h_1,h_2)=(x_2,(x_1\wedge \neg
x_1)\vee \neg x_2)=(x_2,\neg x_2)$ with wiring diagram in Figure
\ref{figure_R} %(see \si for details)
. We can see that the
functionality of $x_3$, that is, ``being $\neg x_1$'', is inherited
to $x_2$ and simplified using $S$.
\end{example}

The Boolean network obtained by eliminating vertices
$x_{i_1},\ldots,x_{i_k}$ is denoted by
$f^{[x_{i_1},\ldots,x_{i_k}]}$ with wiring diagram
$A^{[x_{i_1},\ldots,x_{i_k}]}$. The following proposition states
that the order in which variables are eliminated in not important.

\begin{proposition}
Using the notation of Algorithm R we have the following:
%\begin{enumerate}
Using the notation above we have the following:
$f^{[x_i,x_j]}=f^{[x_j,x_i]}$ and $A^{[x_i,x_j]}=A^{[x_j,x_i]}$

\begin{proof}
Without loss of generality we can assume $x_i=x_1$ and $x_j=x_n$.

Let $g'=(g'_1,\ldots,g'_{n-1})=f^{[x_n]}$,
$g=(g_2,\ldots,g_{n-1})=g'^{[x_1]}=f^{[x_n,x_1]}$,
$h'=(h'_2,\ldots,h'_n)=f^{[x_1]}$,
$h=(h_2,\ldots,h_{n-1})=h'^{[x_n]}=f^{[x_1,x_n]}$. We consider 4
cases.\\

Case 1. $x_n$ and $x_1$ do not depend on each other.

Then, since $f_1$ does not depend on $x_n$ (nor itself), we have
$f_1=f_1(x_2,\ldots,x_{n-1})$ and since $f_n$ does not depend on
$x_1$ (nor itself) we have $f_n=f_n(x_2,\ldots,x_{n-1})$.

Then, $g'_1=f_1(x_1,x_2,\ldots,x_{n-1})$ and
$g'_i=f_i(x_1,x_2,\ldots,x_{n-1},f_n(x_2,\ldots,x_{n-1}))$ for
$i=2,\ldots,n-1$. Then, for $i=2,\ldots,n-1$ we have
\begin{align*}
    g_i &=g'_i(g'_1,x_2,\ldots,x_{n-1})\\
        &=f_i(g'_1,x_2,\ldots,x_{n-1},f_n(x_2,\ldots,x_{n-1}))\\
        &=f_i(f_1(x_2,\ldots,x_{n-1}),x_2,\ldots,x_{n-1},f_n(x_2,\ldots,x_{n-1}))
\end{align*}

Similarly, it follows that $h_i=
f_i(f_1(x_2,\ldots,x_{n-1}),x_2,\ldots,x_{n-1},f_n(x_2,\ldots,x_{n-1}))$
for $i=2,\ldots,n-1$. Hence, $g_i=h_i$ for $i=2,\ldots,n-1$ and
$g=h$.\\

Case 2. $x_n$ depends on $x_1$ but $x_1$ does not depend on $x_n$.

Then, since $f_n$ depends on $x_1$ (and not on $x_n$), we have
$f_n=f_n(x_1,\ldots,x_{n-1})$ and since $f_1$ does not depend on
$x_n$ (nor $x_1$), we have $f_1=f_1(x_2,\ldots,x_{n-1})$.

Then, $g'_1=f_1(x_2,\ldots,x_{n-1})$ and
$g'_i=f_i(x_1,\ldots,x_{n-1},f_n(x_1,\ldots,x_{n-1}))$ for
$i=2,\ldots,n-1$. Then, for $i=2,\ldots,n-1$ we have
\begin{align*}
g_i&=g'_i(g'_1,x_2,\ldots,x_{n-1})\\
&=f_i(g'_1,x_2,\ldots,x_{n-1},f_n(g'_1,x_2,\ldots,x_{n-1}))\\
&=f_i(f_1(x_2,\ldots,x_{n-1}),x_2,\ldots,x_{n-1},f_n(f_1(x_2,\ldots,x_{n-1}),x_2,\ldots,x_{n-1}))
\end{align*}

On the other hand,
$h'_i=f_i(f_1(x_2,\ldots,x_{n-1}),x_2,\ldots,x_n)$ for
$i=2,\ldots,n-1$ and
$h'_n=f_n(f_1(x_2,\ldots,x_{n-1}),x_2,\ldots,x_{n-1})$. Then, for
$i=2,\ldots,n-1$ we have
\begin{align*}
 h_i&=h'_i(x_2,\ldots,x_{n-1},h'_n)\\
&=f_i(f_1(x_2,\ldots,x_{n-1}),x_2,\ldots,x_{n-1},h'_n)\\
&=f_i(f_1(x_2,\ldots,x_{n-1}),x_2,\ldots,x_{n-1},f_n(f_1(x_2,\ldots,x_{n-1}),x_2,\ldots,x_{n-1}))
\end{align*}

Hence, $g_i=h_i$ for $i=2,\ldots,n-1$ and $g=h$.\\

Case 3. $x_1$ depends on $x_n$ but $x_n$ does not depend on $x_1$.
It is analogous to Case 2.\\

Case 4. $x_1$ and $x_n$ depend on each other.

Then, since $f_n$ depends on $x_1$ (and not on $x_n$), we have
$f_n=f_n(x_1,\ldots,x_{n-1})$ and since $f_1$ depends on $x_n$ (and
not on $x_1$), we have $f_1=f_1(x_2,\ldots,x_n)$.

Then, $g'_1=f_1(x_2,\ldots,x_{n-1},f_n(x_1,\ldots,x_{n-1}))$ and
$g'_i=f_i(x_1,\ldots,x_{n-1},f_n(x_1,\ldots,x_{n-1}))$ for
$i=2,\ldots,n-1$. Then, for $i=2,\ldots,n-1$ we have
\begin{align*}
g_i&=g'_i(g'_1,\ldots,x_{n-1})\\
&=f_i(g'_1,x_2,\ldots,x_{n-1},f_n(g'_1,\ldots,x_{n-1}))\\
&=f_i(f_1(x_2,\ldots,x_{n-1},f_n(x_1,\ldots,x_{n-1})),x_2,\ldots,x_{n-1},f_n(f_1(x_2,\ldots,x_{n-1},f_n(x_1,\ldots,x_{n-1})),x_2,\ldots,x_{n-1}))
\end{align*}
On the other hand,
$h'_n=f_n(f_1(x_2,\ldots,x_n),x_2,\ldots,x_{n-1})$ and
$h'_i=f_i(f_1(x_2,\ldots,x_n),x_2,\ldots,x_n)$ for $i=2,\ldots,n-1$.
Then, for $i=2,\ldots,n-1$ we have
\begin{align*}
h_i&=h'_i(x_2,\ldots,x_{n-1},h'_n)\\
&=f_i(f_1(x_2,\ldots,h'_n),x_2,\ldots,x_{n-1},h'_n)\\
&=f_i(f_1(x_2,\ldots,x_{n-1},f_n(f_1(x_2,\ldots,x_n),x_2,\ldots,x_{n-1})),x_2,\ldots,x_{n-1},f_n(f_1(x_2,\ldots,x_n),x_2,\ldots,x_{n-1}))
\end{align*}
Notice that $f_1(x_2,\ldots,x_{n-1},f_n(x_1,\ldots,x_{n-1}))$ does
not depend on $x_1$ and
$f_n(f_1(x_2,\ldots,x_n),x_2,\ldots,x_{n-1})$ does not depend on
$x_n$ ($g=g'^{[x_1]}$ and $h=h'^{[x_n]}$ would be undefined
otherwise). Then, for $i=2,\ldots,n-1$
\begin{align*}
g_i&=f_i(f_1(x_2,\ldots,x_{n-1},f_n(x_1,\ldots,x_{n-1})),x_2,\ldots,x_{n-1},f_n(f_1(x_2,\ldots,x_{n-1},f_n(x_1,\ldots,x_{n-1})),x_2,\ldots,x_{n-1}))\\
&=f_i(f_1(x_2,\ldots,x_{n-1},f_n(x_1,\ldots,x_{n-1})),x_2,\ldots,x_{n-1},f_n(f_1(x_2,\ldots,x_{n-1},x_n),x_2,\ldots,x_{n-1}))\\
&=f_i(f_1(x_2,\ldots,x_{n-1},f_n(f_1(x_2,\ldots,x_n),\ldots,x_{n-1})),x_2,\ldots,x_{n-1},f_n(f_1(x_2,\ldots,x_{n-1},x_n),x_2,\ldots,x_{n-1}))\\
&=h_i
\end{align*}
Hence, $g=h$

\end{proof}
\end{proposition}

The Boolean network obtained by eliminating all variables that can
be eliminated (variables that do not have a self loop) is denoted by
$f^R$ with wiring diagram $A^R$. From Proposition [], it follows
that $f^R$ and $A^R$ are independent of the order chosen to
eliminate vertices (but they do depend on the choice of variables to
be eliminated). Also, it may be the case that $f^R$ and $A^R$ are
empty; in the case they are not empty, each vertex has a self loop.

\begin{example}
\label{example-Th} Consider the Boolean network $f$ defined by:
$$
\begin{array}{ll}
    f_1=x_9\vee x_{11} & f_7=x_4\\
    f_2=\neg x_7\wedge x_{12}  & f_8=x_5\\
    f_3=0          & f_9=x_6\wedge \neg x_{12} \\
    f_4=x_1\wedge \neg x_{10} & f_{10}=x_7\vee x_{11}\\
    f_5=x_2\wedge \neg x_{10} & f_{11}=(x_7\vee x_{11})\wedge \neg x_{12}\\
    f_6=x_3\wedge \neg x_8     & f_{12}=x_8\wedge \neg x_{11}\\
\end{array}
$$

This Boolean network corresponds to the logical model for
Th-lymphocyte differentiation presented in \cite{Remy}. It turns out
that we can eliminate the variables
$x_1,x_2,x_3,x_4,x_5,x_6,x_7,x_8,x_9,x_{10}$. Then
$f^R=h=(h_{11},h_{12})=(x_{11}\wedge \neg x_{12},\neg x_{11}\wedge
x_{12})$ %(see \si for details)
. Notice that each vertex in the the
wiring diagram of the reduced network has a self loop.

\end{example}

%%%%%%%%%%%%%%%%%%%%%%%%%%%%%%%%%%%%%
\section{Properties of the Reduction Method}
\label{sec-pro}
%%%%%%%%%%%%%%%%%%%%%%%%%%%%%%%%%%%%%

%%%%%%%%%%%%%%%%%%%%%%%%%%%%%%%%%%%%%
\subsection{Reduction Method and Dynamical Properties}
\label{sec-pro-dyn}
%%%%%%%%%%%%%%%%%%%%%%%%%%%%%%%%%%%%%

We now show that the original and reduced network share important
dynamical properties. The next theorem states that the reduction
method does not create nor destroy steady states.

\begin{theorem}
\label{main_th} Let $f$ be a Boolean network and
$g=f^{[x_{i_1},x_{i_2},\ldots,x_{i_k}]}$ with $k<n$. Consider the
projection $\pi:\{0,1\}^n\rightarrow \{0,1\}^{n-k}$ defined by\\
$\pi(x_1,\ldots,x_n)=(x_1,\ldots,x_{i_1-1},\check{x}_{i_1},x_{i_1+1},\ldots,x_{i_k-1},\check{x}_{i_k},x_{i_k+1}\ldots,x_n)$.
Then, $\pi$ defines a one to one correspondence between the set of
steady states of $f$ and the set of steady states of $g$.

\begin{proof}
We only need to prove the theorem for $k=1$; the general case
follows by induction. Without loss of generality we can assume that
$g=(g_1,\ldots,g_{n-1})=f^{[x_n]}$.

If $z=(z_1,\ldots,z_n)$ is a steady state of $f$, that is, $f(z)=z$,
or $f_i(z)=z_i$ for $i=1,\ldots,n$, we want to show that
$\pi(z)=(z_1,\ldots,z_{n-1})$ is a steady state of $g$, that is,
$g(\pi(z))=\pi(z)$ or $g_i(\pi(z))=z_i$ for $i=1,\ldots,n-1$. On the
other hand, if $\widetilde{z}=(z_1,\ldots,z_{n-1})$ is a steady
state of $g$, that is, $g_i(\widetilde{z})=z_i$ for
$i=1,\ldots,n-1$, we want to show that there is a unique steady
state of $f$, $z$, such that $\pi(z)=\widetilde{z}$.

Without loss of generality, suppose $x_n$ depends on $x_i$ for
$i=1,\ldots,x_r$; and $x_i$ for $i=s,\ldots,n-1$ are the variables
that depend on $x_n$. Notice that it may be the case that $r=0$ or
$s=n$; in that case $x_n$ would depend on no variables or no
variable would depend on $x_n$. Then, $f_n=f_n(x_1,\ldots,x_r)$;
also, for $i=s,\ldots,n-1$ we have
$f_i=f_i(x_{i1},x_{i2},\ldots,x_n)$. Then, for $i=1,\ldots,s-1$ we
have $g_i(\pi(x))=f_i(x)$ and for $i=s,\ldots,n-1$ we have
$g_i=f_i(x_{i1},x_{i2},\ldots,f_n(x_1,\ldots,x_r))$.

Let $z$ be a steady state of $f$. Then, for $i=1,\ldots,s-1$ we have
$g_i(\pi(z))=f_i(z)=z_i$; for $i=s,\ldots,n-1$,
$g_i(\pi(z))=f_i(z_{i1},z_{i2},\ldots,f_n(z_1,\ldots,z_r))=
f_i(z_{i1},z_{i2},\ldots,z_n)=z_i$. Then, $g(\pi(z))=\pi(z)$.

Let $\widetilde{z}$ be a steady state of $g$. Define
$z=(\widetilde{z},f_n(z_1,\ldots,z_r))$, that is,
$z_n=f_n(z_1,\ldots,z_r)$; we claim that $z$ is a steady state of
$f$. Notice that $\pi(z)=\widetilde{z}$. For $i=1,\ldots,s-1$ we
have that $f_i(z)=g_i(\pi(z))=g_i(\widetilde{z})=z_i$; also, for
$i=s,\ldots,n-1$ we have that
$f_i(z)=f_i(z_{i1},z_{i2},\ldots,z_n)=f_i(z_{i1},z_{i2},\ldots,f_n(z_1,\ldots,z_r))
=g_i(\pi(z))=z_i$; also, $f_n(z)=f_n(z_1,\ldots,z_r)=z_n$. This
shows that $z=(\widetilde{z},f_n(z_1,\ldots,z_r))$ is a steady state
of $f$. We now show that $z$ is unique. Let $z'=(z'_1,\ldots,z'_n)$
be another steady state of $f$ such that $\pi(z')=\widetilde{z}$; it
follows that $z'=(\widetilde{z},z'_n)$. Since
$z'_n=f_n(z')=f_n(z'_1,\ldots,z'_r)=f_n(z_1,\ldots,z_r)=z_n$, then,
$z=(\widetilde{z},z_n)=z'$. Hence, the steady state is unique.
\end{proof}
\end{theorem}

\begin{corollary}
Let $f$ be a Boolean network and $g=f^R$ not empty. Then, there is a
one to one correspondence between the set of steady states of $f$
and the set of steady states of $g$.
\end{corollary}

\begin{corollary}
Let $f$ be a Boolean network and $g=f^R$. If $g$ is empty, then $f$
has a unique steady state.
\begin{proof}
Suppose $g$ is empty. Without loss of generality, suppose $x_1$ is
the last variable to be eliminated. That is,
$h=f^{[x_2,\ldots,x_n]}$ and $g=h^{[x_1]}$. Since the wiring diagram
of $h:\{0,1\}\rightarrow \{0,1\}$ has only the vertex $x_1$ and it
cannot have a self loop, it follows that $h$ cannot be the Boolean
function $h(x_1)=x_1$ or $h(x_1)=\neg x_1$. Then, $h$ has to be the
Boolean function $h(x_1)=0$ or $h(x_1)=1$ and hence it has a single
steady state ($x_1=0$ or $x_1=1$, respectively).
\end{proof}
\end{corollary}

\begin{corollary}
Let $f$ be a Boolean network and
$g=f^{[x_{i_1},x_{i_2},\ldots,x_{i_k}]}$ with $k<n$. Then, $f$ is a
oscillatory system (the only attractors it has are periodic orbits)
if and only if $g$ is a oscillatory system.
\begin{proof}
$f$ is an oscillatory system if and only if $f$ does not have any
steady state if and only if $g$ does not have any steady state if
and only if $g$ is an oscillatory system.
\end{proof}
\end{corollary}

%%%%%%%%%%%%%%%%%%%%%%%%%%%%%%%%%%%%%
\subsection{Reduction Method and Topological Properties}
\label{sec-pro-top}
%%%%%%%%%%%%%%%%%%%%%%%%%%%%%%%%%%%%%

We now show that the original and reduced network share topological
properties. The next theorem states that the reduction method does
not create new paths nor it changes their signs.

\begin{theorem}
Let $f$ be a Boolean network and
$g=f^{[x_{i_1},x_{i_2},\ldots,x_{i_k}]}$ with $k<n$. Then, if there
is path from $y$ to $z$ in the wiring diagram of $g$, there is also
a path from $y$ to $z$ in the wiring diagram of $f$ (or
equivalently, if there is no path from $y$ to $z$ in the wiring
diagram of $f$, there is no path from $y$ to $z$ in the wiring
diagram of $g$). Furthermore, if all paths involved in the reduction
from $y$ to $z$ in the wiring diagram of $f$ are positive
(negative), the corresponding paths from $y$ to $z$ in the wiring
diagram of $g$ are positive (negative).

\begin{proof}
We only need to prove the theorem for $k=1$; the general case
follows by induction. Without loss of generality we can assume that
$g=(g_1,\ldots,g_{n-1})=f^{[x_n]}$. Suppose there is a path from $y$
to $z$ in the wiring diagram of $g$. Without loss of generality we
can suppose the path is $y=x_1\rightarrow x_2 \rightarrow \ldots
\rightarrow x_{r-1}\rightarrow x_r=z$. It follows that $g_i$ depends
on $x_{i-1}$ for $i=2,\ldots,r$.

We claim that there is a path from $x_1$ to $x_2$ in the wiring
diagram of $f$. If $x_2$ depends on $x_n$,
$g_2=f_2(x_{j_1},\ldots,x_{j_t},f_n)$. Then, since $g_2$ depends on
$x_1$, it follows that $x_1$ is one of the $x_{j_l}$'s or $f_n$
depends on $x_1$; then, it follows that there is a path from $x_1$
to $x_2$ in the wiring diagram of $f$. If, on the other hand, $x_2$
does not depend on $x_n$, $g_2=f_2(x_{j_1},\ldots,x_{j_t})$. Then,
since $g_2$ depends on $x_1$, it follows that $x_1$ is one of the
$x_{j_l}$'s; then, there is a path from $x_1$ to $x_2$ in the wiring
diagram of $f$. Similarly, it follows that there is a path from
$x_{i-1}$ to $x_i$ for $i=2,\ldots,r$ in the wiring diagram of $f$
and hence, there is path from $y=x_1$ to $z=x_r$ in the wiring
diagram of $f$.

Now, to conclude the proof of the theorem it is enough to show that
if the paths $x_{n-2}\rightarrow x_{n-1}$ and $x_{n-2}\rightarrow
x_n\rightarrow x_{n-1}$ in the wiring diagram of
$f=(f_1,\ldots,f_n)$ are positive, so is the path $x_1\rightarrow
x_3$ in the wiring diagram of $g=(g_1,\ldots,g_{n-1})$. Since
$f_n(x)=f_n(\ldots,x_{n-2})$ and
$f_{n-1}(x)=f_{n-1}(\ldots,x_{n-2},x_n)$, then
$g_{n-1}(\ldots,x_{n-2})=f_{n-1}(\ldots,x_{n-2},f_n(\ldots,x_{n-2}))$.
To show that the path $x_{n-2}\rightarrow x_{n-1}$ in the wiring
diagram of $g$ is positive, we need to show that
$g_{n-1}(\ldots,0)\leq g_{n-1}(\ldots,1)$. Since the paths
$x_{n-2}\rightarrow x_{n-1}$ and $x_{n-2}\rightarrow x_n\rightarrow
x_{n-1}$ in the wiring diagram of $f$ are positive, then
$f_{n-1}(\ldots,0,f_n(\ldots,0))\leq
f_{n-1}(\ldots,0,f_n(\ldots,1))\leq
f_{n-1}(\ldots,1,f_n(\ldots,1))$. Hence, $g_{n-1}(\ldots,0)\leq
g_{n-1}(\ldots,1)$. Then, the last part of the theorem follows by
induction.
\end{proof}
\end{theorem}

\begin{corollary}
Let $f$ be a Boolean network and $g=f^R$. Then, if there is a path
from $y$ to $z$ in the wiring diagram of $g$, there is also a path
from $y$ to $z$ in the wiring diagram of $f$.
\end{corollary}

\begin{corollary}
Let $f$ be a Boolean network and
$g=f^{[x_{i_1},x_{i_2},\ldots,x_{i_k}]}$ with $k<n$. Then, if there
is a feedback loop at $y$ in the wiring diagram of $g$, there is
also a feedback loop at $y$ in the wiring diagram of $f$.
\end{corollary}

\begin{corollary}
Let $f$ be a Boolean network and $g=f^R$. Then, if there is a self
loop at $y$ in the wiring diagram of $g$, there is also a feedback
loop at $y$ in the wiring diagram of $f$.
\end{corollary}

\begin{figure}[here]
\centerline{ \hbox{ \framebox{
\includegraphics[width=0.65\textwidth]{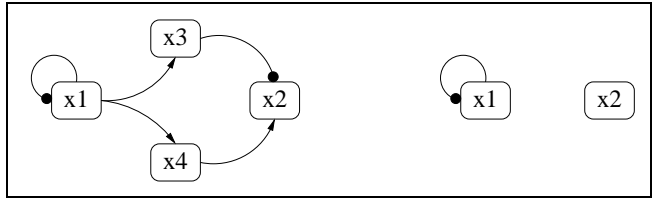}}}
 }\caption{Wiring diagram of the full (left) and reduced (right) network for Example \ref{example_nfpath}.}
 \label{figure_nfpath}
\end{figure}

\begin{example}
\label{example_nfpath} The next example shows how the reduction
method can allow to detect nonfunctional paths.
Consider the Boolean network:\\
$f=(f_1,f_2,f_3,f_4)=(\neg x_1,x_3\vee \neg x_4,x_1,x_1)$ with
wiring diagram shown in Figure \ref{figure_nfpath}. If we reduce the
network by deleting vertices $x_3,x_4$ we obtain
$f^{[x_3,x_4]}=g=(g_1,g_2)=(\neg x_1,1)$. We can clearly see that
the reduced network does not have any path from $x_1$ to $x_2$.
Also, the reduced network does not have any steady state; hence, by
Theorem \ref{main_th}, $f$ does not have steady states.
\end{example}

\begin{figure}[here]
\centerline{ \hbox{ \framebox{
\includegraphics[width=0.55\textwidth]{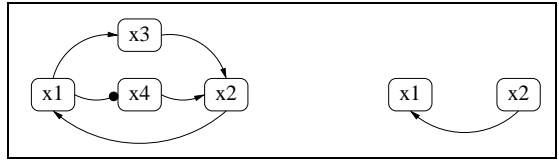}}}
 }\caption{Wiring diagram of the full (left) and reduced (right) network for Example \ref{example_nfloop}.}
 \label{figure_nfloop}
\end{figure}

\begin{example}
\label{example_nfloop} The next example shows how the reduction
method can allow to detect nonfunctional feedback loops.
Consider the Boolean network:\\
$f=(f_1,f_2,f_3,f_4)=(x_2,x_3\vee x_4,x_1,\neg x_1)$ with wiring
diagram shown in Figure \ref{figure_nfloop}. We can see that there
are two feedback loops (one positive and one negative). If we reduce
the network by deleting vertices $x_3,x_4$ we obtain
$f^{[x_3,x_4]}=g=(g_1,g_2)=(x_2,1)$. We can clearly see that the
reduced network does not have any feedback loop. Also, it is easy to
see that $g$ has a unique steady state, (1,1); hence, by Theorem
\ref{main_th} $f$ has a unique steady state.
\end{example}

%%%%%%%%%%%%%%%%%%%%%%%%%%%%%%%%%%%%%
\section{Application}
\label{sec-app}
%%%%%%%%%%%%%%%%%%%%%%%%%%%%%%%%%%%%%

We now consider the Boolean model presented in \cite{Remy}. It is a
small model for Th-lymphocyte differentiation. Its wiring diagram is
given in Figure \ref{figure_Th}. The variables and Boolean functions
of the model are
given as follows:\\

\begin{center}
\begin{tabular}{|l|l|}
  \hline
   variable & Boolean function\\
  \hline
  $x_1$=INF-$\gamma$ &  $f_1=x_9\vee x_{11}$ \\
  \hline
  $x_2$=IL-4 & $f_2=\neg x_7\wedge x_{12}$ \\
  \hline
  $x_3$=IL-12 &  $f_3=0$ \\
  \hline
  $x_4$=IFN-$\gamma$ R &   $f_4=x_1\wedge \neg x_{10}$ \\
  \hline
  $x_5$=IL-4R &   $f_5=x_2\wedge \neg x_{10}$ \\
  \hline
  $x_6$=IL-12R &   $f_6=x_3\wedge \neg x_8$ \\
  \hline
  $x_7$=STAT1 &   $f_7=x_4$ \\
  \hline
  $x_8$=STAT6 &  $f_8=x_5$ \\
  \hline
  $x_9$=STAT4 &  $f_9=x_6\wedge \neg x_{12}$ \\
  \hline
  $x_{10}$=SOCS1 & $f_{10}=x_7\vee x_{11}$ \\
  \hline
  $x_{11}$=T-bet & $f_{11}=(x_7\vee x_{11})\wedge \neg x_{12}$ \\
  \hline
  $x_{12}$=GATA-3 & $f_{12}=x_8\wedge \neg x_{11}$ \\
  \hline
\end{tabular}
\end{center}

\begin{figure}[here]
\centerline{ \hbox{ \framebox{
\includegraphics[width=0.7\textwidth]{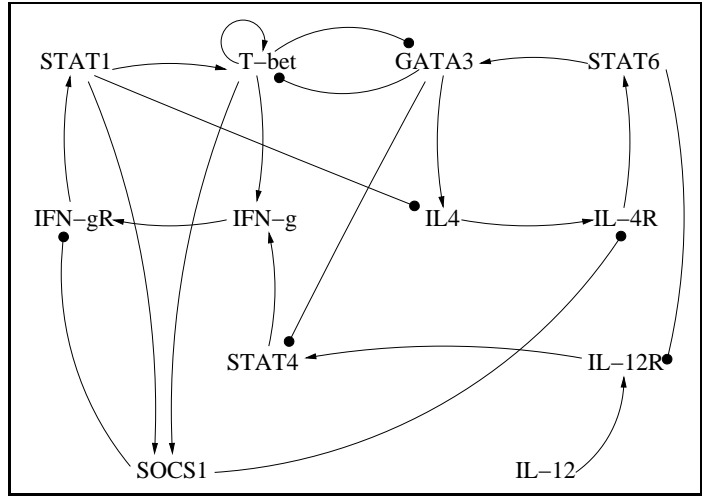}}}
 }\caption{Wiring diagram for the Th-Lymphocyte differentiation network.}
 \label{figure_Th}
\end{figure}

Notice that this Boolean network has $2^{12}=4096$ states.

In Example \ref{example-Th} we observed that the reduction method
gives the following reduced model with wiring diagram given in
Figure \ref{figure_Th}:\\

\begin{center}
\begin{tabular}{|l|l|}
  \hline
   variable & Boolean function\\
  \hline
  $x_{11}$=T-bet &  $h_{11}=x_{11}\wedge \neg x_{12}$ \\
  \hline
  $x_{12}$=GATA-3 & $h_{12}=\neg x_{11}\wedge x_{12}$ \\
  \hline
\end{tabular}
\end{center}

\begin{figure}[here]
\centerline{ \hbox{ \framebox{
\includegraphics[width=0.3\textwidth]{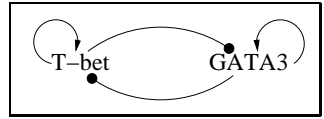}}}
 }\caption{Wiring diagram for the reduced network of the Th-Lymphocyte differentiation network.}
 \label{figure_Th}
\end{figure}

Notice that the reduced network has only $2^2=4$ states which is
about 0.1\% of the number of states of the full network. It is easy
to see that there are 3 steady states:
$(\textbf{0},\textbf{0}),(\textbf{0},\textbf{1}),(\textbf{1},\textbf{0})$.
Furthermore, we can explain the nature of the steady states: if
$x_{12}=$GATA-3=0, then $x_{11}=$T-bet can be 0 or 1  for
$(x_{11},x_{12})$ to be a steady state; on the other hand, if
$x_{12}=$GATA-3=1, then $x_{11}=$T-bet must be 0. Since the reduced
network has 3 steady states, by Theorem \ref{main_th}, the larger
network has also 3 steady states:
$s_1=(0,0,0,0,0,0,0,0,0,0,\textbf{0},\textbf{0}),
s_2=(0,1,0,0,1,0,0,1,0,0,\textbf{0},\textbf{1}),
s_3=(1,0,0,0,0,0,0,0,0,1,\textbf{1},\textbf{0})$. In the reduced
network, the existence of the self loops at GATA3, T-bet and the
positive feedback loop T-bet$\rightleftarrows$ GATA3 suggests that
they are the key to the dynamical properties; this in turn suggests
that the corresponding feedback loops in the larger network are
determining factors in the dynamics.

Now, let us see how the reduced network can help us understand the
larger network. For the reduced network, it is not difficult to see
that deleting one or both of the loops at GATA3 or T-bet results in
the loss of the steady state (0,0). On the other hand, deleting one
of the edges T-bet$\rightarrow$ GATA-3, GATA-3$\rightarrow$ T-bet
does not change the steady states; however, if we delete both edges,
a fourth steady state (1,1) is created. It is important to notice
that this information can be easily obtained. We can expect that the
larger network has similar properties; to check this we study the
effect of deleting edges: deleting the loop at T-bet and other edges
so that we do not have a feedback loop at T-bet, or any edge in the
feedback loop [IL-4,IL-4R, STAT6,GATA-3] results in the loss of the
steady state $s_1$ that corresponds to (T-bet,GATA3)=(0,0). On the
other hand, deleting one of the edges T-bet$\rightarrow$ GATA-3,
GATA-3$\rightarrow$ T-bet does not change the steady states;
however, deleting both edges and other edges so that we do not have
a path from T-bet to GATA3 and GATA3 to T-bet, results in the
creation of a fourth steady state corresponding to
(T-bet,GATA3)=(1,1) %(see \si for details)
. All these properties of
the larger network are consistent with those of the reduced network
that only has 0.1\% of the number of states. In summary, the
reduction method generated a small network that allowed to easily
study the existence and type of steady states and the role of the
feedback loops in the dynamics.

%%%%%%%%%%%%%%%%%%%%%%%%%%%%%%%%%%%%%
\section{Discussion}
\label{sec-disc}
%%%%%%%%%%%%%%%%%%%%%%%%%%%%%%%%%%%%%

Boolean networks have been successfully used in modeling; they
provide a theoretical framework that allows for simulation, control
theory and reverse engineering. Since their analysis is not a
trivial task many mathematical and computational tools have been
developed.

In this paper we have proposed a reduction method that although
simple, it can make the analysis of Boolean networks easier. In
particular, we applied the reduction method to analyze the steady
states of a Boolean model for the Th-lymphocyte differentiation. The
reduction method was not only able to make the analysis of steady
states easier but was also able to explain the role of feedback
loops.

Future work is to study how the reduction method can help in the
analysis of limit cycles of a Boolean network. Also, another future
project is the generalization of the reduction method to general
finite dynamical systems.

\bibliographystyle{plain}
\bibliography{bib}

\end{document}